\documentclass{article}
\title{Untitled}
\author{shamit kachru}

\usepackage{hyperref}

\usepackage{cite}
\usepackage{amsmath}
\usepackage{amssymb}
\usepackage{amsthm}

\usepackage{graphicx,color}

\begin{document}

\centerline{\large{Black Holes and Hurwitz Class Numbers}}
\bigskip
\bigskip
\centerline{Shamit Kachru$^{a,}\footnote{Corresponding author}$, Arnav Tripathy$^b$}
\bigskip
\bigskip
\centerline{$^a$Stanford Institute for Theoretical Physics}
\centerline{Stanford University, Palo Alto, CA 94305, USA}
\centerline{Email: skachru@stanford.edu}
\medskip
\centerline{$^b$Department of Mathematics, Harvard University}
\centerline{Cambridge, MA 02138, USA}
\centerline{Email: tripathy@math.harvard.edu}
\bigskip
\centerline{Submitted: March 12, 2017}
\bigskip
\begin{abstract}
We define a natural counting function for BPS black holes in $K3 \times T^2$ compactification of type II string theory, and observe that it is given by a weight 3/2 mock modular form discovered by Zagier.
This hints at tantalizing relations connecting black holes, string theory, and number theory.

\end{abstract}
\bigskip
\bigskip
\centerline
{\it Essay written for the Gravity Research Foundation 2017 Awards for Essays on Gravitation.}

\newpage

\section{Introduction}

The mathematical foundations of quantum gravity in general, and string theory in particular, remain incomplete.  A framework as
compelling and predictive as Riemannian geometry for the mathematics underlying string theory would certainly help provide a precise
formulation of the theory.  While ties between string theory and algebraic geometry emerged in the 1990s, more intriguing
still are the hints of connections with number theory.  In recent years, indications that deep aspects of the theory of automorphic forms
should underlie string compactification and black hole physics in string theory have started to emerge.  

\medskip
One example involves
the appearance of mock modular forms, discovered by Ramanujan in 1920 but not properly understood until early in this century,
as counting functions for black holes in string theory \cite{DMZ}.  This begins to realize Dyson's dream:

\medskip
\noindent
{\it ``The mock theta-functions give us tantalizing hints of a grand synthesis still to be discovered...My dream is that
I will live to see the day when our young physicists, struggling to bring the predictions of superstring theory into correspondence
with the facts of nature, will be led to enlarge their analytic machinery to include mock theta-functions...''}

\smallskip
\centerline{\it{- Freeman Dyson}\cite{Dyson}}

\medskip
\noindent
This essay will provide an example where a particular (famous) mock modular form mysteriously encodes the black hole physics of
one of the simplest string compactifications.

\medskip
Type II string compactification on a Calabi-Yau manifold $X$ results in a 4d ${\cal N}=2$ supersymmetric low-energy
effective theory.  Such theories admit BPS black holes with various electric and magnetic charges.
The attractor mechanism of Ferrara, Kallosh, and Strominger gives rise to a dynamical system which governs the flow
of (vector multiplet) moduli from their values at infinity in ${\mathbb R^4}$ to the horizon of a BPS black hole with given charges \cite{FKS}.
The values the vector multiplet moduli attain at the horizon are known as `attractor points' in the moduli space.  

\medskip
As the attractor mechanism singles out special points in Calabi-Yau moduli space based on integer data (the choice of an integral cohomology class which specifies the black hole charges), it is natural to hope that the set of attractor points has some
special mathematical significance.  Interesting conjectures in this regard have been put forward by Moore in 
\cite{Mooreone,Mooretwo}.  

\medskip
For $X = K3 \times T^2$, the theory has an enhanced ${\cal N}=4$ supersymmetry, and it is possible to solve for the complete set of attractor points \cite{Mooreone}.  We demonstrate here
that a suitable generating function, summing over all attractor black holes, gives rise to a famous weight
3/2 mock modular form discovered by Zagier \cite{Zagier}.

\section{Attractors on $K3 \times T^2$}

We consider type IIB string compactification on $K3 \times T^2$.  
The BPS black holes arise from D3-branes wrapping three-cycles in the $K3 \times T^2$.  Equivalently, the charges are given by elements of $H^3(K3 \times T^2, {\mathbb Z})$.  Since we have
$$H^3(K3 \times T^2, {\mathbb Z}) \simeq H^2(K3,{\mathbb Z}) \oplus H^2(K3, {\mathbb Z})~,$$
with a map given by considering branes in the first (second) summand to be wrapping $a$ vs $b$ cycles in the $T^2$,
we can say that the electric charges are given by (say) branes wrapping the $a$-cycle in $T^2$, while the magnetic charges are given by branes wrapping the $b$-cycle.

\medskip
As discussed in \cite{Mooreone, Mooretwo}, the conditions for attractors are as follows.  Consider the Neron-Severi lattice of the K3 -- the lattice spanned by (co)homology classes of algebraic curves.  The transcendental lattice is its orthogonal complement in 
$H^2(K3,{\mathbb Z})$.  Then the K3s which satisfy the attractor equations are precisely those for which
the rank of the Neron-Severi lattice is maximal - i.e., it has rank 20.  Such K3s are called `singular' by mathematicians, though they are not singular in the usual sense (as opposed to K3 surfaces with A-D-E singularities, which also play an important role in string theory).

\medskip
A theorem of Shioda and Inose \cite{Shioda} classifies such `singular' K3 surfaces.  Their theorem demonstrates (through an explicit construction) that singular K3 surfaces are in one-to-one correspondence with
PSL(2,${\mathbb Z}$) equivalence classes of binary quadratic forms.  One can think of this quadratic form as the one canonically associated to the transcendental lattice of the singular K3 surface.

\medskip
Specifically, given electric and magnetic charges $q,p$, the attractive K3 surface is the one associated by Shioda-Inose to the quadratic form
$$2 Q_{p,q} = 
\begin{pmatrix} p^2&-p\cdot q\\ -p \cdot q& q^2 \end{pmatrix}
$$

\medskip
\noindent
Furthermore, the complex structure modulus of the $T^2$ factor in the $K3 \times T^2$ attractor geometry is fixed to 
$$\tau(p,q) = {{p \cdot q + i \sqrt{-D_{p,q}} }\over p^2}
$$
where $D_{p,q}$ is the discriminant of the quadratic
form:
$$D(p,q) = (p\cdot q)^2 - p^2 q^2~.$$ (Note that these discriminants $D$ are negative.)

\medskip
The properties of the quadratic form govern the properties of the BPS black hole in space-time.  Most fundamentally, its entropy is determined by the discriminant via the formula
$$S_{\rm black~hole}(p,q) = \sqrt{-D(p,q)}~.$$
But also, the number of other (topological) BPS branes preserving the same supersymmetry as the given attractor black hole -- i.e., Lagrangian cycles $C$ in $K3$ with vanishing
restriction of the K\"ahler form $J\vert_{C} = 0$ -- is given precisely by $-D$.

\medskip
Happily, $D$ is a quartic invariant of the U-duality group $O(22,6;{\mathbb Z})$ enjoyed by this class of compactifications.  While we should count U-duality equivalent black holes only once each, there can of course
be U-duality inequivalent black holes which happen to have the same value of the discriminant.  
The {\bf class number} of a quadratic form counts the number of distinct $SL(2,{\mathbb Z})$ equivalence classes
of forms
that share its value of the discriminant.  So to each $D$ one can associate a class number $Cl(D)$.

\medskip
A slightly more nuanced notion of class number was introduced by Hurwitz.  Here, instead of defining the class number as 

\medskip
$Cl(N) = \# \{ SL(2,{\mathbb Z}) ~\text{equivalence classes of primitive quadratic forms with}$
$~~~~~\text{~discriminant} -N\}~,$

\medskip
we consider the Hurwitz class number

\medskip
$H(N) = \# \{ SL(2,{\mathbb Z})~{\rm equivalence~classes~of~possibly~imprimitive~quadratic}$
$~~~~~{\rm ~forms~with~discriminant~}-N{\rm~weighted~by~the
~inverse~order~of~their~ auto-}\\$
$~~~~~~{\rm morphism~group}\}~.$

\medskip
\noindent
Counting forms weighted by the inverse order of the automorphism group is a natural notion in physics as well.

\section{The black hole counting function}

We can now define a global BPS black hole counting function for $K3 \times T^2$ compactification.  This should be distinguished from the automorphic objects which count BPS states at a given point in moduli space; our object will
instead count objects of a fixed charge which arise {\bf anywhere in moduli space}.

\medskip
Let us set
$${\cal Z}(q) = \sum_{N} H(N)~ q^N~.$$
The sum runs over all possible values of the discriminant of the quadratic form associated to a singular K3 surface.
The weighting by $H(N)$ gives the factor counting the number of U-duality inequivalent black holes corresponding to distinct attractor K3s with the
same discriminant.  

\medskip
Remarkably, the function ${\cal Z}(q)$ has occurred in the mathematics literature.  In fact, it is a famous example
of a mock modular form, first studied by Zagier in 1975 \cite{Zagier}.  It is nicely discussed in an accessible way
in \cite{Gross}.

\medskip
To make the correspondence precise, we should add a constant term $-{1\over 12}$ to ${\cal Z}(q)$.  With this addition,
${\cal Z}$ becomes the holomorphic part of the Eisenstein series of weight 3/2:
$$E_{3/2}(\tau) = -{1\over 12} + \sum_{N} H(N)~ q^N + y^{-1/2} \sum_{n\in{\mathbb Z}} \beta(4\pi n^2y) q^{-n^2}$$
with
$$\beta(x) = {1\over 16\pi} \int_{1}^{\infty} u^{-3/2} e^{-xu} du~,$$
$q=e^{2\pi i \tau}$, and $y = {\rm Im}(\tau)$.
This function exhibits automorphy under the congruence subgroup $\Gamma_0(4) \subset SL(2,{\mathbb Z})$.

\medskip
There are simple connections between the counting of BPS black holes in $K3$ compactification and the study of rational conformal field theories (RCFTs) relevant in studying heterotic strings compactified on a torus.  These are described in \cite{Mooreone,Mooretwo,GukovVafa}.  This gives an alternative interpretation of this mock modular form as a counting of RCFTs in the Narain moduli space arising in heterotic compactification on $T^4$.  Here, $-D$ finds interpretation as governing the number of chiral primaries of the RCFT.

\medskip
It is natural to wonder whether similar counting functions giving global BPS state counts for other compactifications are also related to simple objects in number theory and the theory of automorphic forms.
In fact, as discussed in \cite{KT}, the attractor points on $K3$ moduli space are a very special (maximal) case of the more general phenomenon of ``BPS jumping loci" (which are defined as special cycles in the moduli space where the number of BPS states jumps).  It is also then natural to conjecture that counting functions for other classes of BPS jumping loci in the moduli space of $K3$ compactifications -- which would count higher dimensional special cycles, instead of attractor points -- exhibit interesting number theoretic properties.

\end{document}